\documentclass[aps,prd,twocolumn,amsmath,amssymb,amsfonts,nofootinbib,superscriptaddress,altaffilletter]{revtex4-1}
\usepackage{graphicx}
\usepackage{hyperref}
\hypersetup{
  bookmarksopen=true
}
\usepackage{amssymb}
\usepackage{amsmath}
\usepackage{color}
\usepackage{supertabular}
\usepackage{dcolumn}

\def\EatH{Einstein@Home }

\def\ec{\textrm{~,}}

\def\fdot{f^{(1)}}
\def\GFA{\textrm{GFA}}

\newcommand{\SNR}{\textrm{SNR}}

\def\sci#1#2{#1\times10^{#2}}

\def\RAJ{\textrm{RA}_{\textrm J2000}}
\def\DECJ{\textrm{DEC}_{\textrm J2000}}

\begin{document}

\title{ 
Comprehensive All-sky Search for Periodic Gravitational Waves in the Sixth Science Run LIGO Data\\
}

\author{%
B.~P.~Abbott,$^{1}$  
R.~Abbott,$^{1}$  
T.~D.~Abbott,$^{2}$  
M.~R.~Abernathy,$^{3}$  
F.~Acernese,$^{4,5}$ 
K.~Ackley,$^{6}$  
C.~Adams,$^{7}$  
T.~Adams,$^{8}$ 
P.~Addesso,$^{9}$  
R.~X.~Adhikari,$^{1}$
% ...
\\
\medskip
(LIGO Scientific Collaboration and Virgo Collaboration) 
\\
\medskip
{{}$^{\dag}$Deceased, May 2015. {}$^{\ddag}$Deceased, March 2015. }%
}\noaffiliation
\affiliation {LIGO, California Institute of Technology, Pasadena, CA 91125, USA }
\affiliation {Louisiana State University, Baton Rouge, LA 70803, USA }
\affiliation {American University, Washington, D.C. 20016, USA }
\affiliation {Universit\`a di Salerno, Fisciano, I-84084 Salerno, Italy }
\affiliation {INFN, Sezione di Napoli, Complesso Universitario di Monte S.Angelo, I-80126 Napoli, Italy }
\affiliation {University of Florida, Gainesville, FL 32611, USA }
\affiliation {LIGO Livingston Observatory, Livingston, LA 70754, USA }
\affiliation {Laboratoire d'Annecy-le-Vieux de Physique des Particules (LAPP), Universit\'e Savoie Mont Blanc, CNRS/IN2P3, F-74941 Annecy-le-Vieux, France }
\affiliation {University of Sannio at Benevento, I-82100 Benevento, Italy and INFN, Sezione di Napoli, I-80100 Napoli, Italy }
\affiliation {Albert-Einstein-Institut, Max-Planck-Institut f\"ur Gravi\-ta\-tions\-physik, D-30167 Hannover, Germany }
\affiliation {Nikhef, Science Park, 1098 XG Amsterdam, The Netherlands }
\affiliation {LIGO, Massachusetts Institute of Technology, Cambridge, MA 02139, USA }
\affiliation {Instituto Nacional de Pesquisas Espaciais, 12227-010 S\~{a}o Jos\'{e} dos Campos, S\~{a}o Paulo, Brazil }
\affiliation {INFN, Gran Sasso Science Institute, I-67100 L'Aquila, Italy }
\affiliation {INFN, Sezione di Roma Tor Vergata, I-00133 Roma, Italy }
\affiliation {Inter-University Centre for Astronomy and Astrophysics, Pune 411007, India }
\affiliation {International Centre for Theoretical Sciences, Tata Institute of Fundamental Research, Bangalore 560012, India }
\affiliation {University of Wisconsin-Milwaukee, Milwaukee, WI 53201, USA }
\affiliation {Leibniz Universit\"at Hannover, D-30167 Hannover, Germany }
\affiliation {Universit\`a di Pisa, I-56127 Pisa, Italy }
\affiliation {INFN, Sezione di Pisa, I-56127 Pisa, Italy }
\affiliation {Australian National University, Canberra, Australian Capital Territory 0200, Australia }
\affiliation {The University of Mississippi, University, MS 38677, USA }
\affiliation {California State University Fullerton, Fullerton, CA 92831, USA }
\affiliation {LAL, Univ. Paris-Sud, CNRS/IN2P3, Universit\'e Paris-Saclay, Orsay, France }
\affiliation {Chennai Mathematical Institute, Chennai 603103, India }
\affiliation {Universit\`a di Roma Tor Vergata, I-00133 Roma, Italy }
\affiliation {University of Southampton, Southampton SO17 1BJ, United Kingdom }
\affiliation {Universit\"at Hamburg, D-22761 Hamburg, Germany }
\affiliation {INFN, Sezione di Roma, I-00185 Roma, Italy }
\affiliation {Albert-Einstein-Institut, Max-Planck-Institut f\"ur Gravitations\-physik, D-14476 Potsdam-Golm, Germany }
\affiliation {APC, AstroParticule et Cosmologie, Universit\'e Paris Diderot, CNRS/IN2P3, CEA/Irfu, Observatoire de Paris, Sorbonne Paris Cit\'e, F-75205 Paris Cedex 13, France }
\affiliation {Montana State University, Bozeman, MT 59717, USA }
\affiliation {Universit\`a di Perugia, I-06123 Perugia, Italy }
\affiliation {INFN, Sezione di Perugia, I-06123 Perugia, Italy }
\affiliation {European Gravitational Observatory (EGO), I-56021 Cascina, Pisa, Italy }
\affiliation {Syracuse University, Syracuse, NY 13244, USA }
\affiliation {SUPA, University of Glasgow, Glasgow G12 8QQ, United Kingdom }
\affiliation {LIGO Hanford Observatory, Richland, WA 99352, USA }
\affiliation {Wigner RCP, RMKI, H-1121 Budapest, Konkoly Thege Mikl\'os \'ut 29-33, Hungary }
\affiliation {Columbia University, New York, NY 10027, USA }
\affiliation {Stanford University, Stanford, CA 94305, USA }
\affiliation {Universit\`a di Padova, Dipartimento di Fisica e Astronomia, I-35131 Padova, Italy }
\affiliation {INFN, Sezione di Padova, I-35131 Padova, Italy }
\affiliation {CAMK-PAN, 00-716 Warsaw, Poland }
\affiliation {University of Birmingham, Birmingham B15 2TT, United Kingdom }
\affiliation {Universit\`a degli Studi di Genova, I-16146 Genova, Italy }
\affiliation {INFN, Sezione di Genova, I-16146 Genova, Italy }
\affiliation {RRCAT, Indore MP 452013, India }
\affiliation {Faculty of Physics, Lomonosov Moscow State University, Moscow 119991, Russia }
\affiliation {SUPA, University of the West of Scotland, Paisley PA1 2BE, United Kingdom }
\affiliation {University of Western Australia, Crawley, Western Australia 6009, Australia }
\affiliation {Department of Astrophysics/IMAPP, Radboud University Nijmegen, P.O. Box 9010, 6500 GL Nijmegen, The Netherlands }
\affiliation {Artemis, Universit\'e C\^ote d'Azur, CNRS, Observatoire C\^ote d'Azur, CS 34229, Nice cedex 4, France }
\affiliation {Institut de Physique de Rennes, CNRS, Universit\'e de Rennes 1, F-35042 Rennes, France }
\affiliation {Washington State University, Pullman, WA 99164, USA }
\affiliation {Universit\`a degli Studi di Urbino ``Carlo Bo,'' I-61029 Urbino, Italy }
\affiliation {INFN, Sezione di Firenze, I-50019 Sesto Fiorentino, Firenze, Italy }
\affiliation {University of Oregon, Eugene, OR 97403, USA }
\affiliation {Laboratoire Kastler Brossel, UPMC-Sorbonne Universit\'es, CNRS, ENS-PSL Research University, Coll\`ege de France, F-75005 Paris, France }
\affiliation {Carleton College, Northfield, MN 55057, USA }
\affiliation {Astronomical Observatory Warsaw University, 00-478 Warsaw, Poland }
\affiliation {VU University Amsterdam, 1081 HV Amsterdam, The Netherlands }
\affiliation {University of Maryland, College Park, MD 20742, USA }
\affiliation {Center for Relativistic Astrophysics and School of Physics, Georgia Institute of Technology, Atlanta, GA 30332, USA }
\affiliation {Laboratoire des Mat\'eriaux Avanc\'es (LMA), CNRS/IN2P3, F-69622 Villeurbanne, France }
\affiliation {Universit\'e Claude Bernard Lyon 1, F-69622 Villeurbanne, France }
\affiliation {Universit\`a di Napoli ``Federico II,'' Complesso Universitario di Monte S.Angelo, I-80126 Napoli, Italy }
\affiliation {NASA/Goddard Space Flight Center, Greenbelt, MD 20771, USA }
\affiliation {RESCEU, University of Tokyo, Tokyo, 113-0033, Japan. }
\affiliation {Tsinghua University, Beijing 100084, China }
\affiliation {Texas Tech University, Lubbock, TX 79409, USA }
\affiliation {National Tsing Hua University, Hsinchu City, 30013 Taiwan, Republic of China }
\affiliation {Charles Sturt University, Wagga Wagga, New South Wales 2678, Australia }
\affiliation {West Virginia University, Morgantown, WV 26506, USA }
\affiliation {University of Chicago, Chicago, IL 60637, USA }
\affiliation {Caltech CaRT, Pasadena, CA 91125, USA }
\affiliation {Korea Institute of Science and Technology Information, Daejeon 305-806, Korea }
\affiliation {Universit\`a di Roma ``La Sapienza,'' I-00185 Roma, Italy }
\affiliation {University of Brussels, Brussels 1050, Belgium }
\affiliation {Sonoma State University, Rohnert Park, CA 94928, USA }
\affiliation {Center for Interdisciplinary Exploration \& Research in Astrophysics (CIERA), Northwestern University, Evanston, IL 60208, USA }
\affiliation {University of Minnesota, Minneapolis, MN 55455, USA }
\affiliation {The University of Melbourne, Parkville, Victoria 3010, Australia }
\affiliation {Institute for Plasma Research, Bhat, Gandhinagar 382428, India }
\affiliation {The University of Sheffield, Sheffield S10 2TN, United Kingdom }
\affiliation {The University of Texas Rio Grande Valley, Brownsville, TX 78520, USA }
\affiliation {Universit\`a di Trento, Dipartimento di Fisica, I-38123 Povo, Trento, Italy }
\affiliation {INFN, Trento Institute for Fundamental Physics and Applications, I-38123 Povo, Trento, Italy }
\affiliation {The Pennsylvania State University, University Park, PA 16802, USA }
\affiliation {Cardiff University, Cardiff CF24 3AA, United Kingdom }
\affiliation {Montclair State University, Montclair, NJ 07043, USA }
\affiliation {MTA E\"otv\"os University, ``Lendulet'' Astrophysics Research Group, Budapest 1117, Hungary }
\affiliation {National Astronomical Observatory of Japan, 2-21-1 Osawa, Mitaka, Tokyo 181-8588, Japan }
\affiliation {School of Mathematics, University of Edinburgh, Edinburgh EH9 3FD, United Kingdom }
\affiliation {Indian Institute of Technology, Gandhinagar Ahmedabad Gujarat 382424, India }
\affiliation {University of Szeged, D\'om t\'er 9, Szeged 6720, Hungary }
\affiliation {Embry-Riddle Aeronautical University, Prescott, AZ 86301, USA }
\affiliation {Tata Institute of Fundamental Research, Mumbai 400005, India }
\affiliation {INAF, Osservatorio Astronomico di Capodimonte, I-80131, Napoli, Italy }
\affiliation {University of Michigan, Ann Arbor, MI 48109, USA }
\affiliation {Rochester Institute of Technology, Rochester, NY 14623, USA }
\affiliation {NCSA, University of Illinois at Urbana-Champaign, Urbana, Illinois 61801, USA }
\affiliation {Universitat de les Illes Balears, IAC3---IEEC, E-07122 Palma de Mallorca, Spain }
\affiliation {University of Bia{\l }ystok, 15-424 Bia{\l }ystok, Poland }
\affiliation {SUPA, University of Strathclyde, Glasgow G1 1XQ, United Kingdom }
\affiliation {IISER-TVM, CET Campus, Trivandrum Kerala 695016, India }
\affiliation {Canadian Institute for Theoretical Astrophysics, University of Toronto, Toronto, Ontario M5S 3H8, Canada }
\affiliation {Institute of Applied Physics, Nizhny Novgorod, 603950, Russia }
\affiliation {Pusan National University, Busan 609-735, Korea }
\affiliation {Hanyang University, Seoul 133-791, Korea }
\affiliation {University of Adelaide, Adelaide, South Australia 5005, Australia }
\affiliation {NCBJ, 05-400 \'Swierk-Otwock, Poland }
\affiliation {IM-PAN, 00-956 Warsaw, Poland }
\affiliation {Monash University, Victoria 3800, Australia }
\affiliation {Seoul National University, Seoul 151-742, Korea }
\affiliation {The Chinese University of Hong Kong, Shatin, NT, Hong Kong SAR, China }
\affiliation {University of Alabama in Huntsville, Huntsville, AL 35899, USA }
\affiliation {University of Massachusetts-Amherst, Amherst, MA 01003, USA }
\affiliation {ESPCI, CNRS, F-75005 Paris, France }
\affiliation {Universit\`a di Camerino, Dipartimento di Fisica, I-62032 Camerino, Italy }
\affiliation {Southern University and A\&M College, Baton Rouge, LA 70813, USA }
\affiliation {College of William and Mary, Williamsburg, VA 23187, USA }
\affiliation {Instituto de F\'\i sica Te\'orica, University Estadual Paulista/ICTP South American Institute for Fundamental Research, S\~ao Paulo SP 01140-070, Brazil }
\affiliation {University of Cambridge, Cambridge CB2 1TN, United Kingdom }
\affiliation {IISER-Kolkata, Mohanpur, West Bengal 741252, India }
\affiliation {Rutherford Appleton Laboratory, HSIC, Chilton, Didcot, Oxon OX11 0QX, United Kingdom }
\affiliation {Whitman College, 345 Boyer Avenue, Walla Walla, WA 99362 USA }
\affiliation {National Institute for Mathematical Sciences, Daejeon 305-390, Korea }
\affiliation {Universit\'e de Lyon, F-69361 Lyon, France }
\affiliation {Hobart and William Smith Colleges, Geneva, NY 14456, USA }
\affiliation {Janusz Gil Institute of Astronomy, University of Zielona G\'ora, 65-265 Zielona G\'ora, Poland }
\affiliation {King's College London, University of London, London WC2R 2LS, United Kingdom }
\affiliation {Andrews University, Berrien Springs, MI 49104, USA }
\affiliation {Universit\`a di Siena, I-53100 Siena, Italy }
\affiliation {Trinity University, San Antonio, TX 78212, USA }
\affiliation {University of Washington, Seattle, WA 98195, USA }
\affiliation {Kenyon College, Gambier, OH 43022, USA }
\affiliation {Abilene Christian University, Abilene, TX 79699, USA }

\begin{abstract}
We report on a comprehensive all-sky search for periodic gravitational waves in the frequency band 100-1500\,Hz and with a frequency time derivative in the range of $\sci{[-1.18, +1.00]}{-8}$\,Hz/s. Such a signal could be produced by
a nearby spinning and slightly non-axisymmetric isolated neutron star in our galaxy.
This search uses the data from the Initial LIGO sixth science run and covers a larger parameter space with respect to any past search.
 A {\em Loosely Coherent}  detection pipeline was applied to follow up weak outliers in both Gaussian ($95$\% recovery rate) and non-Gaussian ($75$\% recovery rate) bands.
No gravitational wave signals were observed, and upper limits were placed on their strength. Our smallest upper limit on worst-case (linearly polarized) strain amplitude $h_0$ is  $\sci{9.7}{-25}$ near 169\,Hz, while at the high end of our frequency range we achieve 
a worst-case upper limit of 
$\sci{5.5}{-24}$. Both cases refer to all sky locations and entire range of frequency derivative values. 
\end{abstract}

%
%
\maketitle

\section{Introduction}
\label{sec:introduction}

In this paper we report the results of a comprehensive all-sky search for continuous, nearly monochromatic gravitational waves in data from LIGO's sixth science (S6) run. The search covered frequencies from 100\,Hz through 1500\,Hz and frequency derivatives from $-\sci{1.18}{-8}$\,Hz/s through $\sci{1.00}{-8}$\,Hz/s.

A number of searches for periodic gravitational waves have been carried out previously in LIGO data~\cite{S4IncoherentPaper, EarlyS5Paper, FullS5Semicoherent, orionspur, S2TDPaper, S3S4TDPaper, S2FstatPaper, Crab, pulsars3, CasA}, including coherent searches for gravitational radiation from known radio and X-ray pulsars. An \EatH search running on the BOINC infrastructure \cite{BOINC} has performed blind all-sky searches on S4 and S5 data~\cite{S4EH, S5EH, FullS5EH}.

The results in this paper were produced with the PowerFlux search program. It was first described in \cite{S4IncoherentPaper} together with two other semi-coherent search pipelines (Hough, Stackslide). The sensitivities of all three methods were compared, with PowerFlux showing better results in
frequency bands lacking severe spectral artifacts.
A subsequent article~\cite{FullS5Semicoherent} based on the data from the S5 run featured improved upper limits and a systematic outlier follow-up search based on the {\em Loosely Coherent} algorithm~\cite{loosely_coherent}.

The analysis of the data set from the sixth science run described in this paper has several distinguishing features from previously published results:
\begin{itemize}
 \item A number of upgrades to the detector were made in order to field-test the technology for Advanced LIGO interferometers. This resulted in a factor of about two improvement in intrinsic noise level at high frequencies compared to previously published results \cite{FullS5Semicoherent}. 
 \item The higher sensitivity allowed us to use less data while still improving upper limits in high frequency bands by $25$\% over previously published results. This smaller dataset allowed covering larger parameter space, and comprehensive exploration of high frequency data.
 \item This search improved on previous analyses by partitioning the data in $\approx 1$ month chunks and looking for signals in any contiguous sequence of these chunks. This enables detections of signals that conform to ideal signal model over only part of the data. Such signals could arise because of a glitch, or because of influence of a long-period companion object.
 \item The upgrades to the detector, while improving sensitivity on average, introduced a large number of detector artifacts, with $20$\% of frequency range contaminated by non-Gaussian noise. We addressed this issue by developing a new {\em Universal statistic} \cite{universal_statistics} that provides correct upper limits regardless of the noise distribution of the underlying data, while still showing close to optimal performance for Gaussian data.

\end{itemize}

We have observed no evidence of gravitational radiation and have established the most sensitive upper limits to date in the frequency band 100-1500\,Hz. 
Our smallest 95\% confidence level upper limit on worst-case (linearly polarized) strain amplitude $h_0$ is  $\sci{9.7}{-25}$ near 169\,Hz, while at the high end of our frequency range we achieve 
a worst-case upper limit of 
$\sci{5.5}{-24}$. Both cases refer to all sky locations and entire range of frequency derivative values.

\begin{figure*}[htbp]
\begin{center}
  \includegraphics[width=7.2in]{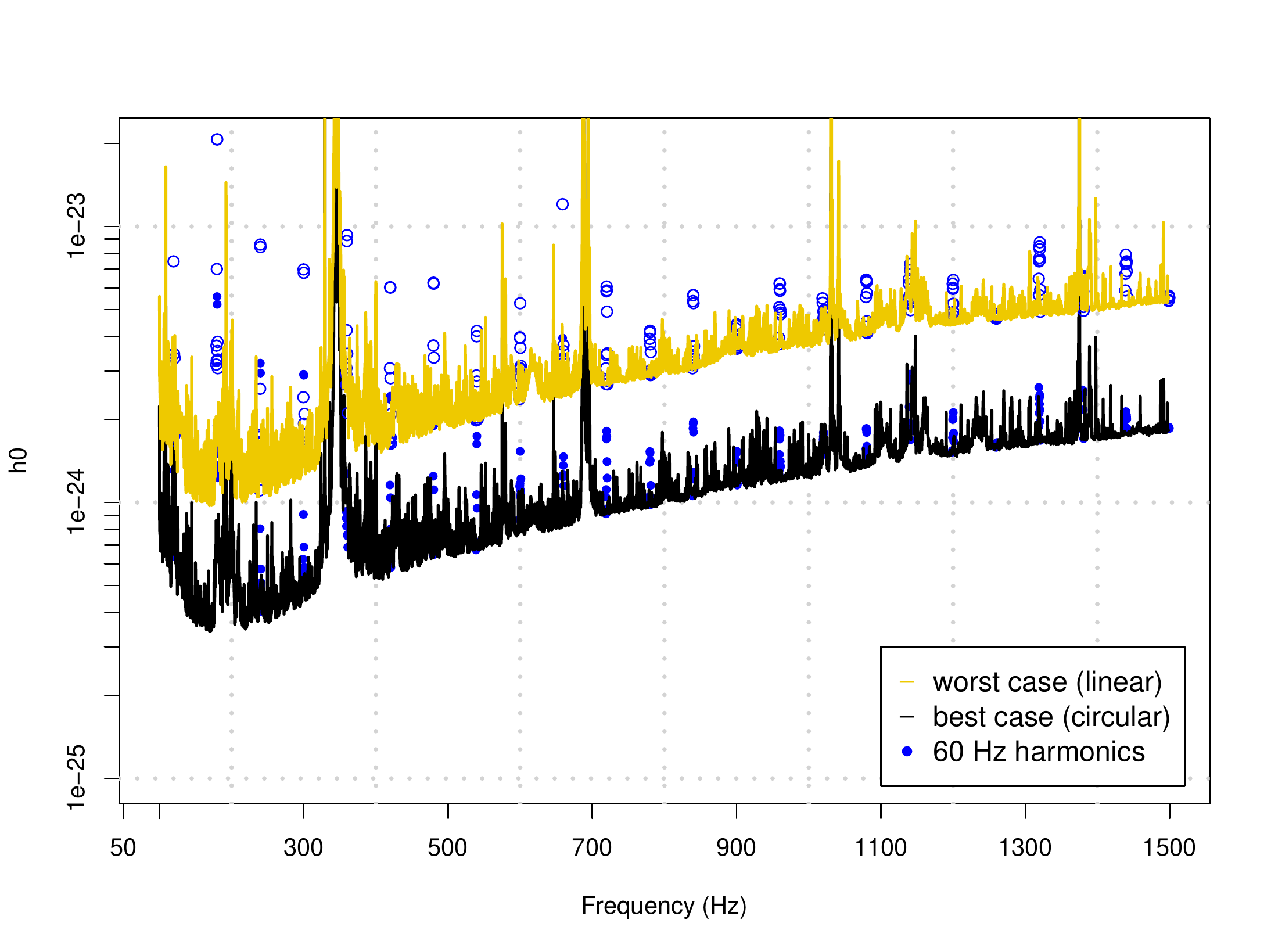}

 \caption{S6 upper limits. The upper (yellow) curve shows worst-case (linearly polarized) $95$\%~CL upper limits in analyzed 0.25\,Hz bands (see Table~\ref{tab:excluded_ul_bands} for list of excluded bands). The lower (grey) curve shows upper limits assuming a circularly polarized source.
 The values of solid points and circles mark frequencies within
$1.25$\,Hz of $60$\,Hz power line harmonics for circularly (solid points) and
linearly (open circles) polarized sources. The data for this plot can be found in \cite{data}. (color online)}
\label{fig:full_s6_upper_limits}
\end{center}
\end{figure*}

\section{LIGO interferometers and S6 science run}

The LIGO gravitational wave network consists of two observatories, one in Hanford, Washington and the other in Livingston, Louisiana, separated by a 3000\,km baseline. During the S6 run each site housed one suspended interferometer with 4\,km long arms. 

While the sixth science run spanned a $\approx 15$\,months  period of data acquisition, this analysis used only data from GPS 951534120 (2010  Mar 02 03:01:45 UTC) through GPS 971619922 (2010 Oct 20 14:25:07 UTC), for which strain sensitivity was best. Since interferometers sporadically fall out of operation (``lose lock'') due to environmental or instrumental disturbances or for scheduled maintenance periods, the dataset was not contiguous. The Hanford interferometer H1 had a duty factor of 53\%, while the Livingston interferometer L1 had a duty factor of 51\%. The strain sensitivity was not uniform, exhibiting a $\sim 50$\% daily variation from anthropogenic activity as well as gradual improvement toward the end of the run \cite{LIGO_detector, detchar}.

Non-stationarity of noise was especially severe at frequencies
below $100$~Hz, and since the average detector sensitivity for such frequencies was not significantly
better than that observed in the longer S5 run \cite{FullS5Semicoherent},
this search was restricted to frequencies above $100$~Hz.

A detailed description of the instruments and data can be found in \cite{detchar2}.

\section{The search for continuous gravitational radiation}
\subsection{Overview}

In this paper we assume a classical model of a spinning neutron star with a rotating quadrupole moment that produces circularly polarized gravitational radiation along the rotation axis and linearly polarized radiation in the directions perpendicular to the rotation axis. The linear polarization is the worst case as such signals contribute the smallest amount of power to the detector.

The strain signal template is assumed to be 
\begin{equation}
\begin{array}{l}
h(t)=h_0\left(F_+(t, \alpha_0, \delta_0, \psi)\frac{1+\cos^2(\iota)}{2}\cos(\Phi(t))+\right.\\
\quad\quad\quad \left.\vphantom{\frac{1+\cos^2(\iota)}{2}}+F_\times(t, \alpha_0, \delta_0, \psi)\cos(\iota)\sin(\Phi(t))\right)\ec
\end{array}
\end{equation}

\noindent where $F_+$ and $F_\times$ characterize the detector responses to signals with ``$+$'' and ``$\times$'' 
quadrupolar polarizations \cite{S4IncoherentPaper, EarlyS5Paper, FullS5Semicoherent}, the sky location is described by right ascension $\alpha_0$ and declination $\delta_0$, the inclination of the source rotation axis to the line of sight is denoted $\iota$, and the phase evolution of the signal is given by the formula
\begin{equation}
\label{eqn:phase_evolution}
\Phi(t)=2\pi\left(f_\textrm{source}\cdot (t-t_0)+\fdot\cdot (t-t_0)^2/2\right)+\phi\ec
\end{equation}
with $f_\textrm{source}$ being the source frequency and $\fdot$ denoting the first frequency derivative (which, when negative, is termed the {\em spindown}). 
We use $t$ to denote the time in the Solar System barycenter frame. The initial phase
$\phi$ is computed relative to reference time $t_0$.  When expressed as a function of local time of ground-based detectors the equation \ref{eqn:phase_evolution} acquires  sky-position-dependent Doppler shift terms.
We use $\psi$ to denote the polarization angle of the projected source rotation axis in the sky plane.

The search has two main components. First, the main {\em PowerFlux} algorithm \cite{S4IncoherentPaper, EarlyS5Paper, FullS5Semicoherent, PowerFluxTechNote, PowerFlux2TechNote, PowerFluxPolarizationNote} was run to establish upper limits and produce lists of outliers with signal-to-noise ratio (SNR) greater than 5. Next, the {\em Loosely Coherent} detection pipeline \cite{loosely_coherent, loosely_coherent2, FullS5Semicoherent} was used to reject or confirm collected outliers. 

Both algorithms calculate power for a bank of signal model templates and compute upper limits and signal-to-noise ratios for each template based on comparison to templates with nearby frequencies and the same sky location and spindown. The input time series is broken into $50$\% overlapping $1800$\,s long segments which are Hann windowed and Fourier transformed. The resulting short Fourier transforms (SFTs) are arranged into an input matrix with time and frequency dimensions. The power calculation can be expressed as a bilinear form of the input matrix $\left\{a_{t,f}\right\}$:

\begin{equation}
P[f] = \sum_{t_1, t_2} a_{t_1, f+\delta f(t_1)} a_{t_2, f+\delta f(t_2)}^* K_{t_1, t_2, f}
\end{equation}
Here $\delta f(t)$ denotes the detector frame frequency drift due to the effects from both Doppler shifts and the first frequency derivative. The sum is taken over all times $t$ corresponding to the midpoint of the short Fourier transform time interval. The kernel $K_{t_1, t_2, f}$ includes the contribution of time dependent SFT weights, antenna response, signal polarization parameters and relative phase terms \cite{loosely_coherent, loosely_coherent2}.

The main semi-coherent PowerFlux algorithm uses a kernel with main diagonal terms only and is very fast. The Loosely Coherent algorithms increase coherence time while still allowing for controlled deviation in phase \cite{loosely_coherent}. This is done by more complicated kernels that increase effective coherence length. 

The effective coherence length is captured in a parameter $\delta$,
which describes the amount of phase drift that the kernel allows between SFTs, with $\delta=0$ corresponding to a fully coherent case, and $\delta=2\pi$ corresponding to incoherent power sums.

Depending on the terms used, the data from different interferometers can be combined incoherently (such as in stages 0 and 1, see Table \ref{tab:followup_parameters}) or coherently (as used in stages 2, 3 and 4). The coherent combination is more computationally expensive but provides much better parameter estimation.

The upper limits (Figure~\ref{fig:full_s6_upper_limits})  are reported in terms of the worst-case value of $h_0$ (which applies to linear polarizations with $\iota=\pi/2$) and for the most sensitive circular polarization ($\iota=0$~or~$\pi$).
As described in the previous paper \cite{FullS5Semicoherent}, the pipeline does retain some sensitivity, however, to non-general-relativity GW polarization models, including a longitudinal component, and to slow amplitude evolution.

\begin{table*}[htbp]
\begin{center}
\begin{tabular}{lp{10cm}}\hline
Category & Description \\
\hline \hline
First harmonic of violin modes & 343.25-343.75\,Hz, 347-347.25\,Hz  \\
Second harmonic of violin modes & 686.25-687.5\,Hz \\
Third harmonic of violin modes & 1031.00-1031.25\,Hz \\
\hline
\end{tabular}
\caption[Frequency regions excluded from upper limit analysis]{Frequency regions excluded from upper limit analysis. ``Violin modes'' are resonant vibrations of the wires which suspend the many mirrors of the interferometer.}
\label{tab:excluded_ul_bands}
\end{center}
\end{table*}

The 95\% confidence level upper limits (see Figure~\ref{fig:full_s6_upper_limits}) produced in the first stage are based on the overall noise level and largest outlier in strain found for every template in each 0.25\,Hz band in
the first stage of the pipeline. The $0.25$\,Hz bands are analyzed by separate instances of PowerFlux \cite{FullS5Semicoherent}.
A followup search for detection is carried out for high-SNR outliers found in the first stage. 
Certain frequency ranges (Table~\ref{tab:excluded_ul_bands}) were excluded from the analysis because of gross contamination by detector artifacts.

\subsection{Universal statistics}

The detector sensitivity upgrades introduced many artifacts, so that in $20$\% of the sensitive frequency range the noise follows non-Gaussian distributions which, in addition, are unknown. As the particular non-Gaussian distribution can vary widely depending on particular detector artifacts, the ideal estimate based on full knowledge of the distribution is not usually available. In the previous analysis \cite{S4IncoherentPaper, EarlyS5Paper, FullS5Semicoherent}, the frequency bands where the noise distribution was non-Gaussian were not used to put upper limits. However, in the present case this approach would have resulted in excluding most of the frequency bands below 400\,Hz and many above 400\,Hz; even though the average strain sensitivity in many of these frequency bands is better than in the past.

\begin{figure}[htbp]
\begin{center}
  \includegraphics[width=3.0in]{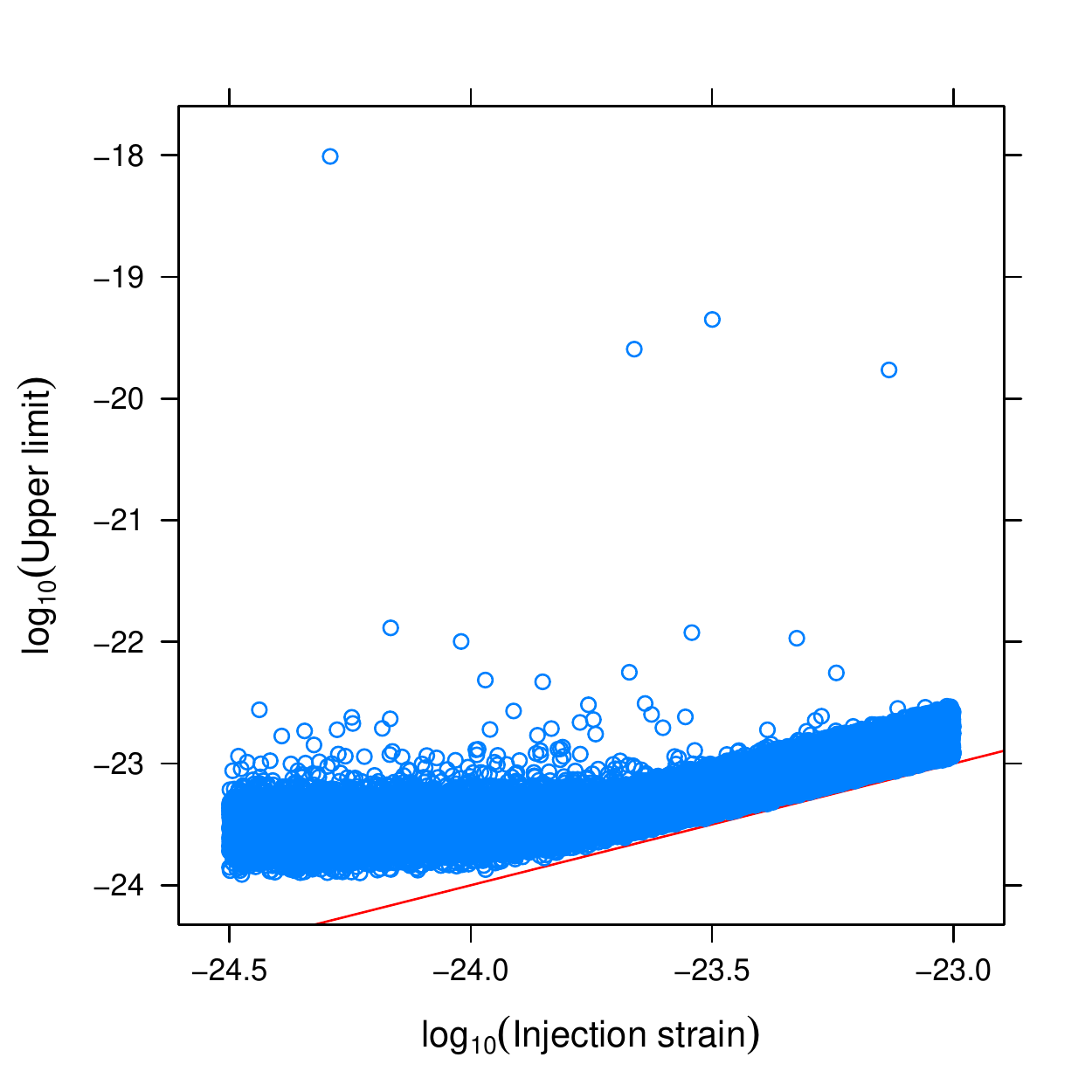}
 \caption[Upper limit versus injected strain]{Upper limit validation. Each point represents a separate injection in the 400-1500\,Hz frequency range. Each established upper limit (vertical axis) is compared against the injected strain value (horizontal axis, red line) (color online).}
\label{fig:ul_vs_strain}
\end{center}
\end{figure}

To make use of the entire spectrum, we used in this work the {\em Universal statistic} algorithm \cite{universal_statistics} for establishing upper limits. The algorithm is derived from the Markov inequality and shares its independence from the underlying noise distribution. It produces upper limits less than $5$\% above optimal in case of Gaussian noise. In non-Gaussian bands it can report values larger than what would be obtained if the distribution were known, but the upper limits are always at least 95\% valid. Figure~\ref{fig:ul_vs_strain} shows results of an injection run performed as described in \cite{FullS5Semicoherent}. Correctly established upper limits are above the red line.

\begin{table*}[htbp]
\begin{center}
\begin{tabular}{llccccc}\hline
Stage & Instrument sum & {Phase coherence} & \multicolumn{1}{c}{Spindown step} & \multicolumn{1}{c}{Sky refinement} & \multicolumn{1}{c}{Frequency refinement} & \multicolumn{1}{c}{SNR increase} \\
 & & \multicolumn{1}{c}{rad} & \multicolumn{1}{c}{Hz/s} &  &  & \multicolumn{1}{c}{\%}\\
\hline \hline \\

  0  & Initial/upper limit semi-coherent & NA  & $\sci{2}{-10}$ & $1$ & $1/2$ & NA \\
  1 & incoherent & $\pi/2$ & $\sci{1.0}{-10}$ & $1/4$ & $1/8$ & 20 \\
  2 & coherent & $\pi/2$ & $\sci{5.0}{-11}$ & $1/4$ & $1/8$ & 0  \\
  3 & coherent & $\pi/4$ & $\sci{2.5}{-11}$ & $1/8$ & $1/16$ & 12 \\
  4 & coherent & $\pi/8$ & $\sci{5.0}{-12}$ & $1/16$ & $1/32$ & 12 \\
 \hline
\end{tabular}
\caption[Analysis pipeline parameters]{Analysis pipeline parameters. Starting with stage 1, all stages used the Loosely Coherent algorithm for demodulation. The sky and frequency refinement parameters are relative to values used in the semicoherent PowerFlux search.}
\label{tab:followup_parameters}
\end{center}
\end{table*}

\subsection{Detection pipeline}

The detection pipeline used in \cite{FullS5Semicoherent} was extended with additional stages (see Table \ref{tab:followup_parameters}) to winnow the larger number of initial outliers, expected because of non-Gaussian artifacts and larger initial search space. This detection pipeline was also used in the search of the Orion spur \cite{orionspur}.

The initial stage (marked 0) scans the entire sky with semi-coherent algorithm that computes weighted sums of powers of $1800$\,s Hann-windowed SFTs. These power sums are then analyzed to identify high-SNR outliers. A separate algorithm uses universal statistics \cite{universal_statistics} to establish upper limits. 

The entire dataset was partitioned into 7 segments of equal length and power sums were produced independently for any contiguous combinations of these stretches. As in \cite{orionspur} the outlier identification was performed independently in each stretch.

High-SNR outliers were subject to a coincidence test.  For each
outlier with $\SNR>7$ in the combined H1 and L1 data, we required there
to be outliers in the individual detector data that had $\SNR>5$,
matching the parameters of the combined-detector outlier within a
distance of $0.03\textrm{\,rad}\cdot 400\textrm{\,Hz}/f$ on the sky, $2$\,mHz in frequency, and
$\sci{3}{-10}$\,Hz/s in spindown.  However, the combined-detector SNR could not
be \emph{lower} than either single-detector SNR.

The identified outliers using combined data are then passed to followup stage using Loosely Coherent algorithm \cite{loosely_coherent} with progressively tighter phase coherence parameters $\delta$, and improved determination of frequency, spindown and sky location.

As the initial stage 0 only sums powers it does not use relative phase between interferometers, which results in some degeneracy between sky position, frequency and spindown. The first Loosely Coherent followup stage also combines
interferometer powers incoherently, but demands greater temporal
coherence (smaller $\delta$) within each interferometer, which should
boost SNR of viable outiers by at least 20\%. Subsequent stages use data coherently providing tighter bounds on outlier location.

The testing of the pipeline was done above $400$\,Hz and included both Gaussian and non-Gaussian bands. We focused on high frequency performance because preliminary S6 data indicated the sensitivity at low frequencies was unlikely to improve over S5 results due to detector artifacts.

\begin{figure}[htbp]
\begin{center}
 \includegraphics[width=3.0in]{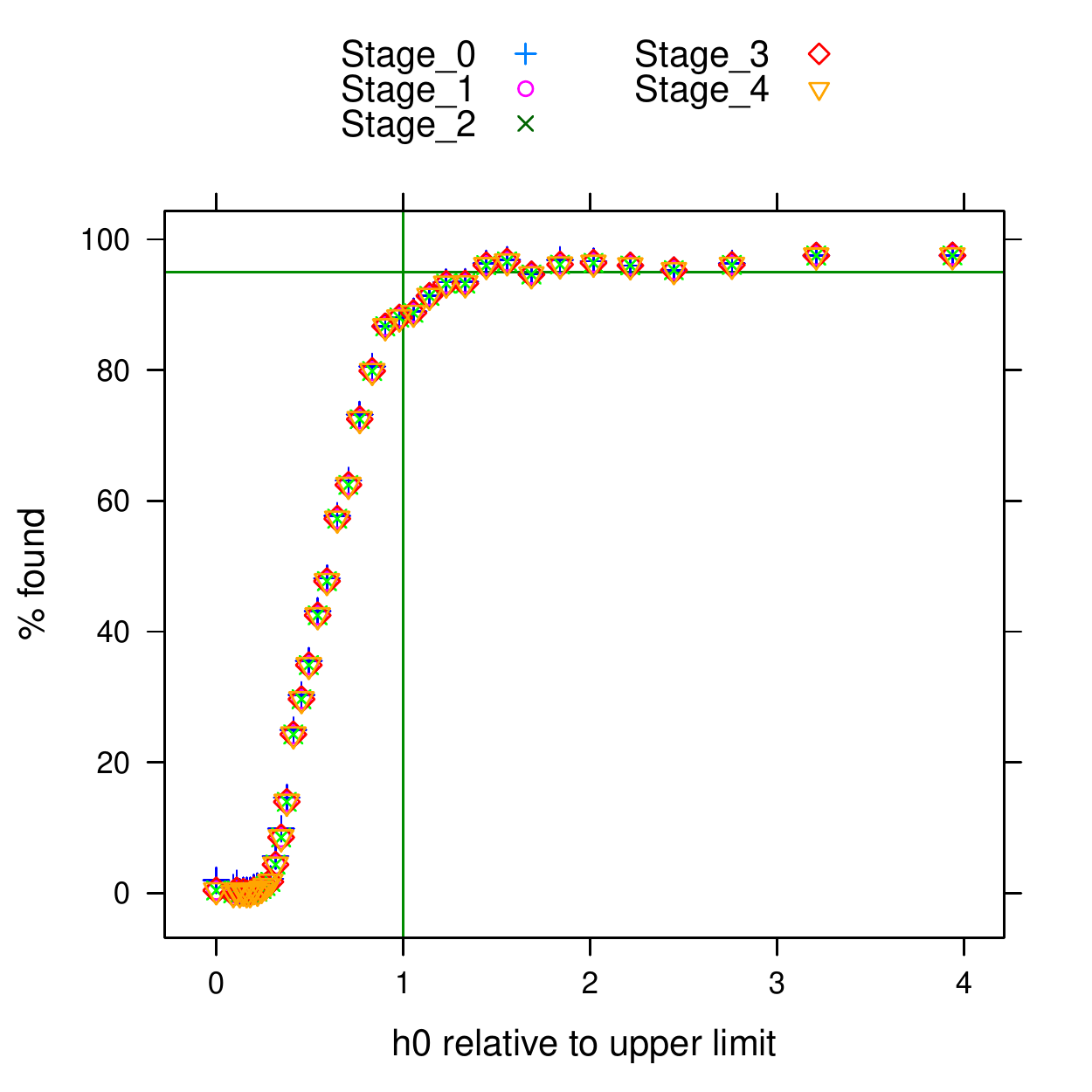}
 \caption[Injection recovery]{Injection recovery in frequency bands above 400\,Hz. The injected strain divided by the upper limit in this band (before injection) is shown on the horizontal axis. The percentage of surviving injections is shown on the vertical axis, with horizontal line drawn at $95$\% level. Stage 0 is the output of the coincidence test after the initial semi-coherent search. (color online). }
\label{fig:injection_recovery}
\end{center}
\end{figure}

\begin{figure}[htbp]
\begin{center}
 \includegraphics[width=3.0in]{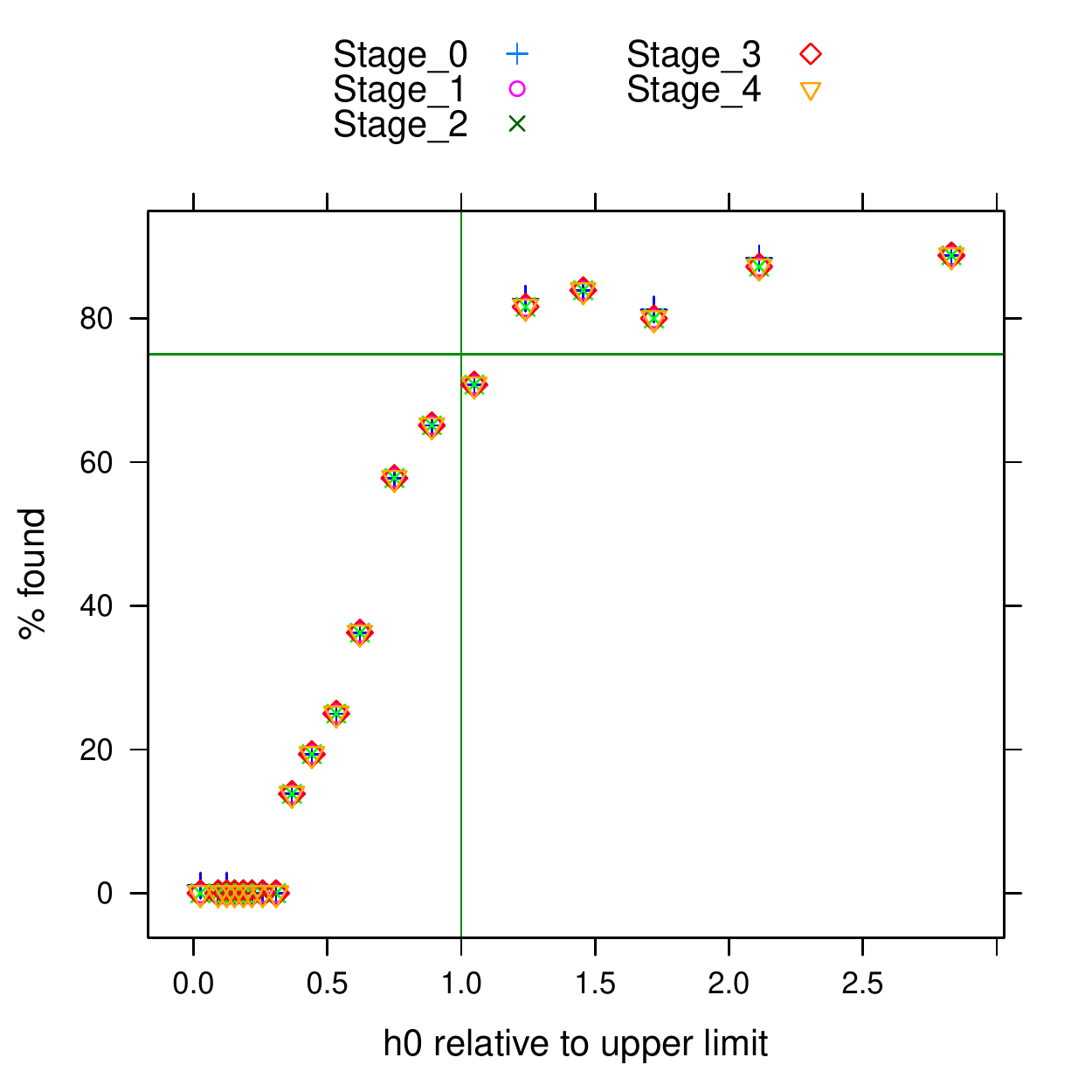}
 \caption[Injection recovery in non-Gaussian bands]{Injection recovery in non-Gaussian bands above 400\,Hz. The injected strain divided by the upper limit in this band (before injection) is shown on the horizontal axis. The percentage of surviving injections is shown on the vertical axis, with horizontal line drawn at $75$\% level. (color online) }
\label{fig:injection_recovery_non_gauss}
\end{center}
\end{figure}

The followup code was tested to recover $95$\% of injections $50$\% above the upper limit level assuming uniform distribution of injection frequency. (Figure \ref{fig:injection_recovery}). 
Recovery of signals injected into frequency bands which exhibits non-Gaussian noise was $75$\% (Figure \ref{fig:injection_recovery_non_gauss}).
Our recovery criterion demanded that an outlier close to the true injection location (within $2$\,mHz in frequency $f$, $\sci{3}{-10}$\,Hz/s in spindown and $12$\,rad$\cdot$Hz$/f$ in sky location) be found and successfully pass  through all stages of the detection pipeline. As each stage of the pipeline only passes outliers with an increase in SNR, this resulted in an outlier that strongly stood out above the background, with good estimates of the parameters of the underlying signal.

It should be noted that the injection recovery curve in Figure 3    
passes slightly below the $95$\% level for $h_0$ equal to the upper limit. 
However, the upper limits are based on power levels measured by stage
0, independent of any follow-up criteria.  That is, we can say with
$95$\% confidence that a signal above the upper limit level is
inconsistent with the observed power, even though such a
(hypothetical) signal might not pass all of our follow-up criteria to
be ``detected''.  The main reason that these injections fail to be
detected is the different sensitivities of the H1 and L1
detectors.  When one interferometer is less sensitive sensitive we can still set a good upper limit, but the initial coincidence criteria requires that an outlier be marginally seen in both interferometers. In the previous analysis \cite{FullS5Semicoherent} the interferometers had similar sensitivity and the curve passed through the intersection of the green lines (horizontal axis value of $1$, vertical axis value of $95$\%).

\subsection{Gaussian false alarm event rate}
\label{sec:GFA}

The computation of the false alarm rate for the outliers passing all stages of the pipeline is complicated by the fact that most outliers are caused by instrumental artifacts for which we do not know the underlying probability distribution. In principle, one could repeat the analysis many times using non-physical frequency shifts (which would exclude picking up a real signal by accident) in order to obtain estimates of false alarm rate, but this approach is very computationally expensive.
Even assuming a perfect Gaussian background, it is difficult to analytically model the pipeline in every detail to obtain an accurate estimate of the false alarm rate, given the gaps in interferometer operations and non-stationary noise.

Instead, following \cite{orionspur}, we compute a figure of merit that overestimates the actual Gaussian false alarm event rate. We simplify the problem by assuming that the entire analysis was carried out with the resolution of the very last stage of follow-up and we are merely triggering on the SNR value of the last stage.
This is extremely conservative as we ignore the consistency requirements that allow the outlier to proceed from one stage of the pipeline to the next; the actual false alarm rate could be lower.

The SNR of each outlier is computed relative to the Loosely Coherent power sum for 501 frequency bins spaced at $1/1800$~Hz intervals (including the outlier) but with all the other signal parameters held constant. The spacing assures that correllations between neighboring sub-bins do not affect the statistics of the power sum. 

To simplify computation we  assume that we are dealing with a simple $\chi^2$ distribution with the number of degrees of freedom given by the timebase divided by the coherence length and multiplied by a conservative duty factor reflecting interferometer uptime and the worst-case weights from linearly-polarized signals.

Thus to find the number $N$ of degrees of freedom we will use the formula

\begin{equation}
\label{N_chi2}
N\approx \frac{\textrm{timebase} \cdot \delta  \cdot \textrm{duty factor}}{ 1800\textrm{~s} \cdot 2\pi }
\end{equation}
with the duty factor taken to be $0.125$ and $\delta$ giving the phase coherence parameter of the Loosely Coherent search. The duty factor was chosen to allow for only $50$\% interferometer uptime and only one quarter of the data receiving high weights from our procedure, which weights the contribution of data inversely as the square of the estimated noise \cite{PowerFluxTechNote, PowerFlux2TechNote}.

Thus we define the outlier figure of merit describing Gaussian false alarm (GFA) event rate as 
\begin{equation}
\label{GFA}
\GFA = K\cdot P_{\chi^2}\left(N+\textrm{SNR}\cdot \sqrt{2 N} ; N\right)
\end{equation}
where $N$ defines the number of degrees of freedom as given by equation \ref{N_chi2}, $P_{\chi^2}(x ; N)$ gives the probability for a $\chi^2$ distribution with $N$ degrees of freedom to exceed $x$, and $K=\sci{1.3}{14}$ is the estimated number of templates.

We point out that the $\GFA$ is overly conservative when applied to frequency bands with Gaussian noise, but is only loosely applicable to bands with detector artifacts, which can affect both the estimate of the number of degrees of freedom of the underlying distribution and the assumption of uncorrelated underlying noise.

\section{Results}
\label{sec:results}

\begin{figure}[htbp]
\begin{center}
  \includegraphics[width=3.0in]{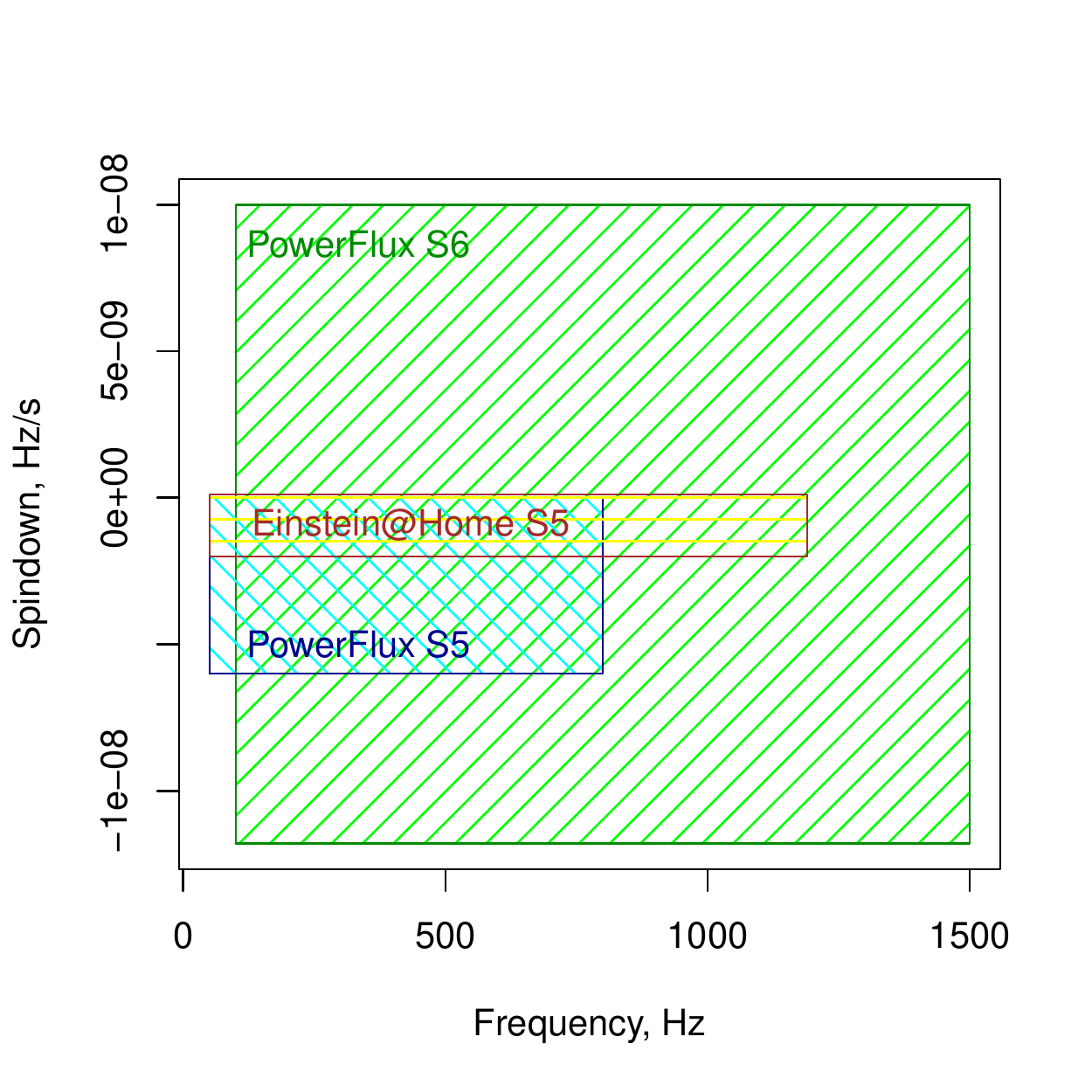}

 \caption{Parameter space covered in the analysis. Einstein@Home searches use longer coherence times than PowerFlux, with better sensitivity to narrow band signals. The results for area marked ``PowerFlux S6'' are reported in this paper. (color online)}
\label{fig:parameter_coverage}
\end{center}
\end{figure}

The PowerFlux algorithm and Loosely Coherent method compute power estimates for gravitational waves in a given frequency band for a fixed set of templates. The template parameters usually include frequency, first frequency derivative and sky location.

Since the search target is a rare monochromatic signal, it would contribute excess power to one of the frequency bins after demodulation. The upper limit on the maximum excess relative to the nearby power values can then be established. For this analysis we use a universal statistic \cite{universal_statistics} that places conservative 95\% confidence level upper limits for an arbitrary statistical distribution of noise power. The universal statistic has been designed to provide close to optimal values in the common case of Gaussian distribution.

The PowerFlux algorithm and Loosely Coherent method have been described in detail in \cite{loosely_coherent, PowerFlux2TechNote, PowerFluxTechNote, PowerFluxPolarizationNote, EarlyS5Paper, S4IncoherentPaper}. 

Most natural sources are expected to have negative first frequency derivative, as the energy lost in gravitational or electromagnetic waves would make the source spin more slowly. The frequency derivative can be positive when the source is affected by a strong slowly-variable Doppler shift, such as due to a long-period orbit.

The large gap in data taking due to installation of Advanced LIGO interferometers provided an opportunity to cover an extended parameter space (Figure  \ref{fig:parameter_coverage}). With respect to previous searches, we have chosen to explore comprehensively both negative and positive frequency derivatives to avoid missing any unexpected sources in our data. 

The upper limits obtained in the search are shown in figure \ref{fig:full_s6_upper_limits}. The numerical data for this plot can be obtained separately \cite{data}. The upper (yellow) curve shows the upper limits for a worst-case (linear) polarizations when the smallest amount of gravitational energy is projected towards Earth. The lower curve shows upper limits for an optimally oriented source. Because of the day-night variability of the interferometer sensitivity due to anthropogenic noise, the linearly polarized sources are more susceptible to detector artifacts, as the detector response to such sources varies with the same period. The neighborhood of 60\,Hz harmonics is shown as circles for worst case upper limits and dots for circular polarization upper limits. Thanks to the use of universal statistic they do represent valid values even if contaminated by human activity.

Each point in figure \ref{fig:full_s6_upper_limits} represents a maximum over the sky: only a small excluded portion of the sky near ecliptic poles that is highly susceptible to detector artifacts, due to stationary frequency evolution produced by the combination of frequency derivative and Doppler shifts. The exclusion procedure is described in \cite{FullS5Semicoherent} and applied to $0.033$\% of the sky over the entire run.

A few frequency bands shown in Table \ref{tab:excluded_ul_bands} were so contaminated that every SFT was vetoed by data conditioning code and the analysis terminated before reaching universal statistic stage. While the universal statistic could have established upper limits with veto turned off, we opted to simply exclude these bands, as the contamination would raise upper limits to be above physically interesting values.

\begin{figure}[htbp]
\includegraphics[width=3in]{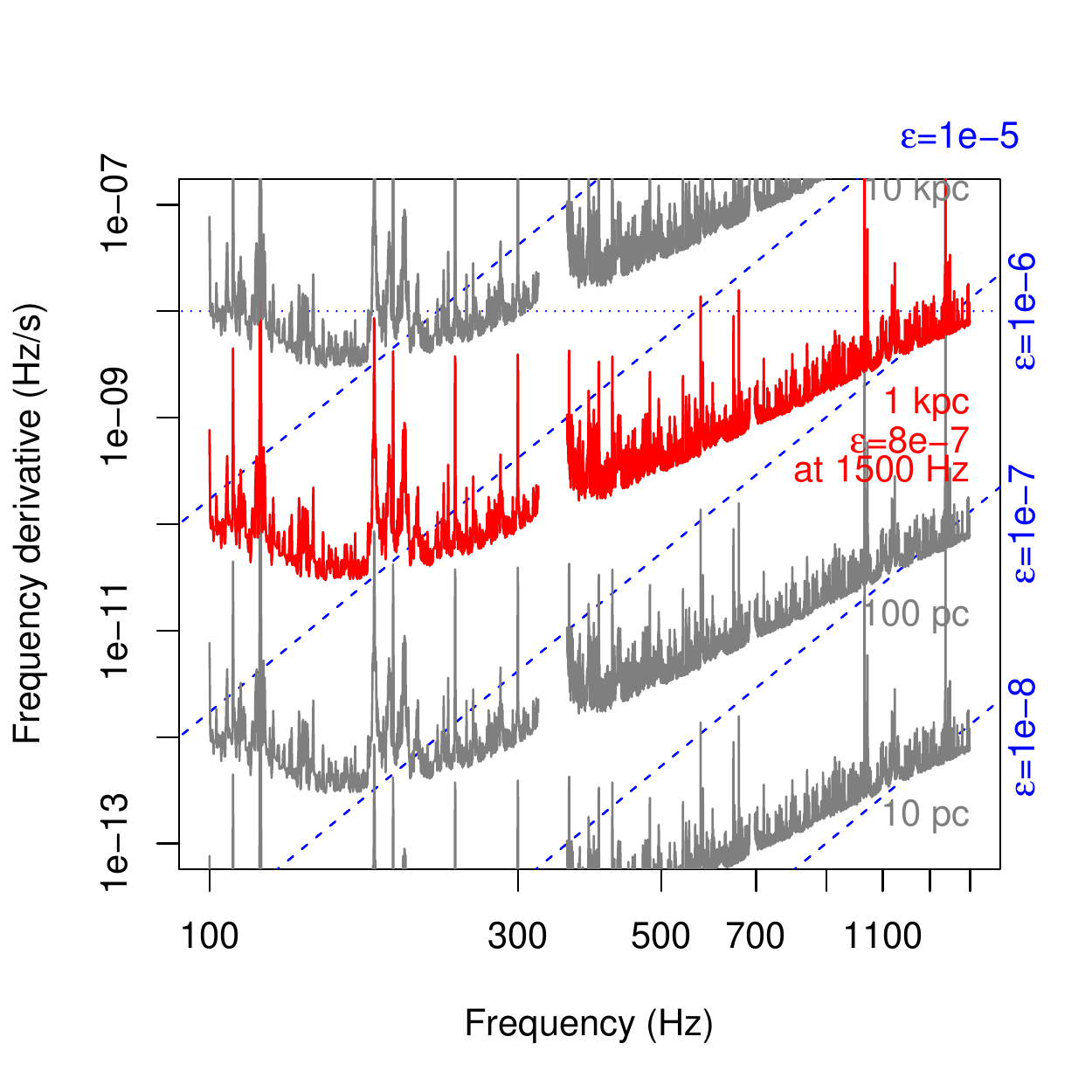}
\caption[Spindown range]{
\label{fig:spindown_range}
Range of the PowerFlux search for neutron stars
spinning down solely due to gravitational radiation.  This is a
superposition of two contour plots.  The grey and red solid lines are contours of the maximum distance at which a neutron
star could be detected as a function of gravitational-wave frequency
$f$ and its derivative $\dot{f}$.  The dashed lines 
are contours of the corresponding ellipticity
$\epsilon(f,\dot{f})$. The fine dotted line marks the maximum spindown searched. Together these quantities tell us the
maximum range of the search in terms of various populations (see text
for details) (color online).}
\end{figure}

If one assumes that the source spindown is solely due to emission of gravitational waves, then it is possible to recast upper limits on source amplitude as a limit on source ellipticity. Figure \ref{fig:spindown_range} shows the reach of our search under different assumptions on source distance. Superimposed are lines corresponding to sources of different ellipticities. 

\begin{table*}[htbp]
\begin{center}
\small
\begin{tabular}{D{.}{.}{2}D{.}{.}{2}D{.}{.}{2}cD{.}{.}{5}D{.}{.}{4}D{.}{.}{4}D{.}{.}{4}l}\hline
\multicolumn{1}{c}{Idx} & \multicolumn{1}{c}{SNR} & \multicolumn{1}{c}{$\log_{10}(\GFA)$} & \multicolumn{1}{c}{Segment} &  \multicolumn{1}{c}{Frequency} & \multicolumn{1}{c}{Spindown} &  \multicolumn{1}{c}{$\RAJ$}  & \multicolumn{1}{c}{$\DECJ$} & Description \\
\multicolumn{1}{c}{}	&  \multicolumn{1}{c}{}	&  \multicolumn{1}{c}{}	&  \multicolumn{1}{c}{}	& \multicolumn{1}{c}{Hz}	&  \multicolumn{1}{c}{nHz/s} & \multicolumn{1}{c}{degrees} & \multicolumn{1}{c}{degrees} & \\
\hline \hline
1 & 3331 & -9360 & $[0,6]$ & 192.49269 & -8.650 & 351.371 & -33.342 & Hardware injection ip8\\
21 & 1329 & -3114 & $[1,5]$ & 108.85717 & -0.000 & 178.417 & -33.400 & Hardware injection ip3,\\ &&&&&&&& Non Gaussian, disturbed H1 spectrum\\
42 & 957 & -2622 & $[0,6]$ & 575.16354 & 0.005 & 215.261 & 3.370 & Hardware injection ip2\\
69 & 112 & -196 & $[0,3]$ & 397.51894 & -0.115 & 271.698 & 67.257 & Non Gaussian, Line in H1, disturbed spectrum in L1\\
72 & 93 & -78 & $[4,4]$ & 1397.76097 & -11.220 & 296.704 & -16.069 & Induced by loud hardware injection ip4,\\ &&&&&&&& Non Gaussian, highly disturbed H1+L1 spectra\\
76 & 82 & -162 & $[0,5]$ & 1145.20043 & 0.400 & 90.936 & -67.610 & Highly disturbed H1 spectrum, stationary line area\\
79 & 64 & -98 & $[1,4]$ & 566.08359 & -4.850 & 91.028 & 86.915 & Line in H1 at 566.085 Hz\\
81 & 54 & -68 & $[2,4]$ & 704.03500 & 4.110 & 117.932 & 50.411 & Disturbed H1 and L1 spectrum\\
82 & 48 & -86 & $[0,6]$ & 1220.74448 & -1.120 & 223.413 & -20.502 & Hardware injection ip7, sloping H1 and L1 spectra\\
83 & 48 & -73 & $[0,4]$ & 140.41014 & -0.010 & 270.298 & 66.821 & Highly disturbed H1 spectrum, stationary line area\\
94 & 36 & -44 & $[0,3]$ & 192.65413 & 9.270 & 145.440 & 10.439 & Induced by loud hardware injection ip8\\
95 & 35 & -28 & $[2,3]$ & 250.01082 & 2.750 & 247.459 & -76.842 & Lines in H1 and L1, Non Gaussian\\
101 & 19 & -13 & $[2,6]$ & 1145.30312 & 8.515 & 196.471 & 33.778 & Highly disturbed H1 spectrum\\
102 & 18 & -12 & $[0,4]$ & 1397.91328 & 1.070 & 42.627 & 32.827 & Induced by loud hardware injection ip4,\\ &&&&&&&& Non Gaussian, highly disturbed H1+L1 spectra\\
103 & 17 & -4 & $[3,4]$ & 1143.41710 & -2.455 & 107.611 & -56.347 & Highly disturbed H1 spectrum\\
107 & 14 & -0 & $[2,3]$ & 451.47993 & -10.880 & 49.317 & 33.890 & Line in H1 at 451.5 Hz\\

\hline
\end{tabular}
\caption[Outliers that passed detection pipeline]{Outliers that passed detection pipeline. Only the highest-SNR outlier is shown for each 0.1\,Hz frequency region. Outliers marked with ``line'' had strong narrowband disturbances identified near the outlier location. Outliers marked as ``non Gaussian'' were identified as having non Gaussian statistics in their power sums, often due to a very steeply sloping spectrum. GFA is the Gaussian false alarm
figure of merit described in Sec. \ref{sec:GFA}. Segment column reports the set of
contiguous segments of the data that produced the outlier, as
described in \ref{sec:results}.
Frequencies are converted to epoch GPS $961577021$.}
\label{tab:outliers}
\end{center}
\end{table*}


\begin{table*}[htbp]
\begin{center}
\begin{tabular}{lD{.}{.}{6}rD{.}{.}{5}D{.}{.}{4}}
\hline
Label & \multicolumn{1}{c}{Frequency} & \multicolumn{1}{c}{Spindown} & \multicolumn{1}{c}{$\RAJ$} & \multicolumn{1}{c}{$\DECJ$} \\
 & \multicolumn{1}{c}{Hz} & \multicolumn{1}{c}{nHz/s} & \multicolumn{1}{c}{degrees} & \multicolumn{1}{c}{degrees} \\
\hline \hline
ip2   &  575.16354  & $\sci{-1.37}{-4}$   &     215.25617   &    3.4440 \\
ip3   &  108.85716 & $\sci{-1.46}{-8}$   &      178.37257  &   -33.4366  \\
ip4   &  1397.831947 & $-25.4$ & 4.88671 &  -12.4666 \\
ip7   &  1220.744496 & $-1.12$   &      223.42562  &   -20.4506  \\
ip8   & 192.492709 & $-0.865$   &       351.38958  &   -33.4185   \\
\hline
\end{tabular}
\caption[Parameters of hardware injections]{Parameters of hardware-injected simulated signals detected by PowerFlux (epoch GPS $961577021$).}
\label{tab:injections}
\end{center}
\end{table*}

The detection pipeline produced 16 outliers (Table \ref{tab:outliers}). Each outlier is identified by a numerical index. We report SNR, decimal logarithm of Gaussian false alarm rate, as well as frequency, spindown and sky location. 

The ``Segment'' column describes the persistence of the outlier through
the data, and specified which contiguous subset of the 7 equal
partitions of the timespan contributed most significantly to the
outlier: see \cite{orionspur} for details.  A continuous signal will normally have
[0,6] in this column (similar contribution from all 7 segments), or on
rare occasions [0,5] or [1,6].  Any other range is indicative of a
non-continuous signal or artifact.

During the S6 run several simulated pulsar signals were injected into the data by applying a small force to the interferometer mirrors. 
Several outliers were due to such hardware injections (Table \ref{tab:injections}). The full list of injections including those too weak to be found by an all-sky search can be found in \cite{SNRpaper}. The hardware injection ip3 was exceptionally strong with a clear signature even in non-Gaussian band. 

The recovery of the injections gives us confidence that no potential signal were missed. Manual followup has shown non-injection outliers to be caused by pronounced detector artifacts.

\section{Conclusions}

We have performed the most sensitive all-sky search to date for continuous gravitational waves 
in the range 100-1500\,Hz. We explored both positive and negative spindowns and placed upper limits on expected and unexpected sources.  At the highest frequencies we are sensitive to neutron stars with an equatorial 
ellipticity as small as $\sci{8}{-7}$ and as far away as $1$\,kpc for favorable spin orientations. The use of a universal statistic allowed us to place upper limits on both Gaussian and non-Gaussian frequency bands. The maximum ellipticity a neutron star can theoretically support is at least $\sci{1}{-5}$ according to \cite{crust_limit, crust_limit2}. Our results exclude such maximally deformed pulsars above $200$\,Hz pulsar rotation frequency ($400$\,Hz gravitational frequency) within $1$\,kpc.

A detection pipeline based on a {\em Loosely Coherent} algorithm was applied to outliers from our search. 
This pipeline was demonstrated to be able to detect simulated signals at the upper limit level for both Gaussian and non-Gaussian bands. Several outliers passed all stages of the coincidence pipeline; their parameters are shown in table \ref{tab:outliers}. However, manual examination revealed no true pulsar signals.
\section{Acknowledgments}

The authors gratefully acknowledge the support of the United States
National Science Foundation (NSF) for the construction and operation of the
LIGO Laboratory and Advanced LIGO as well as the Science and Technology Facilities Council (STFC) of the
United Kingdom, the Max-Planck-Society (MPS), and the State of
Niedersachsen/Germany for support of the construction of Advanced LIGO 
and construction and operation of the GEO600 detector. 
Additional support for Advanced LIGO was provided by the Australian Research Council.
The authors gratefully acknowledge the Italian Istituto Nazionale di Fisica Nucleare (INFN),  
the French Centre National de la Recherche Scientifique (CNRS) and
the Foundation for Fundamental Research on Matter supported by the Netherlands Organisation for Scientific Research, 
for the construction and operation of the Virgo detector
and the creation and support  of the EGO consortium. 
The authors also gratefully acknowledge research support from these agencies as well as by 
the Council of Scientific and Industrial Research of India, 
Department of Science and Technology, India,
Science \& Engineering Research Board (SERB), India,
Ministry of Human Resource Development, India,
the Spanish Ministerio de Econom\'ia y Competitividad,
the Conselleria d'Economia i Competitivitat and Conselleria d'Educaci\'o, Cultura i Universitats of the Govern de les Illes Balears,
the National Science Centre of Poland,
the European Commission,
the Royal Society, 
the Scottish Funding Council, 
the Scottish Universities Physics Alliance, 
the Hungarian Scientific Research Fund (OTKA),
the Lyon Institute of Origins (LIO),
the National Research Foundation of Korea,
Industry Canada and the Province of Ontario through the Ministry of Economic Development and Innovation, 
the Natural Science and Engineering Research Council Canada,
Canadian Institute for Advanced Research,
the Brazilian Ministry of Science, Technology, and Innovation,
Funda\c{c}\~ao de Amparo \`a Pesquisa do Estado de S\~ao Paulo (FAPESP),
Russian Foundation for Basic Research,
the Leverhulme Trust, 
the Research Corporation, 
Ministry of Science and Technology (MOST), Taiwan
and
the Kavli Foundation.
The authors gratefully acknowledge the support of the NSF, STFC, MPS, INFN, CNRS and the
State of Niedersachsen/Germany for provision of computational resources.

This document has been assigned LIGO Laboratory document number \texttt{LIGO-P1500219-v19}.

\newpage

\end{document}